\documentclass[runningheads]{llncs}
\usepackage{graphicx}
\usepackage{lineno,hyperref,color,soul,xcolor}
\hypersetup{
    colorlinks=true,
    linkcolor=blue,
    filecolor=magenta,      
    urlcolor=cyan,
}
 
\begin{document}
\title{Statistical Analysis Driven Optimized Deep Learning System for Intrusion Detection}
\titlerunning{Statistical Analysis Driven Optimized DL System for Intrusion Detection}
\author{Cosimo Ieracitano\inst{1} \and
Ahsan Adeel\inst{2,*} \and
Mandar Gogate\inst{2}\and
Kia Dashtipour\inst{2}\and
Francesco Carlo Morabito\inst{1}\and
Hadi Larijani\inst{3}\and
Ali Raza\inst{4}\and
Amir Hussain\inst{2}}
\authorrunning{C. Ieracitano et al.}%
\institute{DICEAM Department, University Mediterranea of Reggio Calabria, 89124 Reggio Calabria, Italy \\
%\email{\{cosimo.ieracitano,morabito\}@unirc.it}\\
 \and
Department of Computing Science and Mathematics, Faculty of Natural Sciences, University of
Stirling, FK9 4LA Stirling, United Kingdom\\
\and
Department of Communication, Network and Electronic Engineering, Glasgow Caledonian University, G4 0BA Glasgow, United Kingdom\\
%\email{\{cosimo.ieracitano,morabito\}@unirc.it}\\
\and
Department of Networks and Security Rochester, Institute of Applied Technology, Dubai, United Arab Emirates  \\
\email{\inst{*}\ email:ahsan.adeel@stir.ac.uk}}

%\institute{DICEAM Department, University Mediterranea of Reggio Calabria, 89124 Reggio Calabria, Italy\and
%\email{\{cosimo.ieracitano,morabito\}@unirc.it}\\
% \and
%Department of Computing Science and Mathematics, Faculty of Natural Sciences, University of
%Stirling\\
%\email{\{,\}@uni-heidelberg.de}}
%
\maketitle              % typeset the header of the contribution
\begin{abstract}
Attackers have developed ever more sophisticated and intelligent ways to hack information and communication technology (ICT) systems. The extent of damage an individual hacker can carry out upon infiltrating a system is well understood. A potentially catastrophic scenario can be envisaged where a nation-state intercepting encrypted financial data gets hacked. Thus, intelligent cybersecurity systems have become inevitably important for improved protection against malicious threats. However, as malware attacks continue to dramatically increase in volume and complexity, it has become ever more challenging for traditional analytic tools  to detect and mitigate threat. Furthermore, a huge amount of data produced by large networks have made the recognition task even more complicated and challenging. In this work, we propose an innovative statistical analysis driven optimized deep learning system for intrusion detection. The proposed intrusion detection system (IDS) extracts optimized and more correlated features using big data visualization and statistical analysis methods, followed by a deep autoencoder (AE) for potential threat detection. Specifically, a preprocessing module eliminates the outliers and converts categorical variables into one-hot-encoded vectors. The feature extraction module discards features with null values and selects the most significant features as input to the deep autoencoder model (trained in a greedy-wise manner). The NSL-KDD dataset from the Canadian Institute for Cybersecurity is used as a benchmark to evaluate the feasibility and effectiveness of the proposed architecture. Simulation results demonstrate the potential of our proposed IDS system and its  outperformance as compared to existing state-of-the-art methods and recently published novel approaches. Ongoing work includes further optimization and real-time evaluation of our proposed IDS. 
\keywords{Cybersecurity \and Optimized Deep Learning \and Autoencoder \and Big Data Visualization}
\end{abstract}

\section{Introduction}

The heterogeneity of data in modern networks and variety of new protocols have made the intrusion detection ever more complex and challenging. In this context, there is a great deal of interest in developing intelligent, robust and efficient Intrusion Detection Systems (IDS) capable of identifying the potential or unforeseen threat and consequently denying access to the system.
In the literature, traditional machine learning algorithms have been widely employed to develop IDS \cite{tsai2009intrusion}. However, these techniques are based on handcrafted features by expert users and remain deficient  to handle  high-dimensional volume of training data. 
Deep learning (DL) is an advanced machine learning technique that addresses  the limitations of shallow machine learning algorithms \cite{lecun2015deep} by learning feature representation at varying level of granularity directly from raw input data through a deep hierarchical structure. Since DL has shown to achieve human-level performances in several real-world applications  (i.e. health care \cite{gasparini2018information},\cite{morabito2016deep}, sentiment analysis \cite{dashtipour2017persian},\cite{dashtipour2016persent}, saliency detection \cite{wang2018deep},\cite{yan2018unsupervised}) recently, researchers have proposed  
several novel DL driven cybersecurity algorithms \cite{najafabadi2015deep}.
DL driven solutions are capable of efficiently analyzing  big data and identifying  temporal structures in long complex sequences in real-time.

\noindent In this paper, an innovative Statistical Analysis Driven Optimized DL System for Intrusion Detection is proposed. Specifically, a statistical-driven deep autoencoder (AE) is developed to detect normal and abnormal traffic patterns. The proposed framework has been evaluated using the recent NSL-KDD dataset (updated version of the previous KDD Cup 99 (KDD99) dataset \cite{tavallaee2009detailed}). The proposed framework constitutes three main modules as shown in Figure \ref{fig:KDD_method}: \textit{data preprocessing}, \textit{feature extraction}, and \textit{classification}. \textit{Data preprocessing} discards the outliers and converts categorical variables into one-hot-encoded vectors. \textit{Feature extraction} selects the most correlated features and discards features with null values grater than 80\%; For \textit{ classification}, a deep autoencoder and a shallow multilayer perceptron (MLP) classifier are used to classify different categories of the NSL-KDD dataset (Normal, DoS, R2L, Probe). The deep AE and shallow MLP classifiers are compared with four recent models which have been on NSL-KDD dataset. Experimental results (Table \ref{table5}) showed that the deep AE classifier outperformed all other approaches and achieved accuracy up to 87\%.

\noindent The remainder of this paper is organized as follows. Section \ref{sec:Related works}, illustrates previous approaches, especially based on DL architectures trained with NSL-KDD dataset. Section \ref{sec:Methodology} describes the proposed methodology, including NSL-KDD dataset description, data preprocessing, feature extraction and classification. Section \ref{sec:Experimental results} discusses the experimental results. Finally, Section \ref{sec:Conclusion} concludes the paper. 

\section{Related work}
The KDD99 and NSL-KDD datasets have been widely employed as benchmarks to assess the performance and effectiveness of different intrusion detection models.
\noindent Alrawashdeh et al. \cite{alrawashdeh2016toward} developed a deep belief network (DBN) based on Restricted Boltzmann Machine (RBM) modules, followed by a multi-class softmax layer. The model was tested on 10\% of the KDD99 test dataset and achieved a detection accuracy up to 97.9\% with a false alarm rate of 2.47\%.  
%Alom et al. \cite{alom2015intrusion} also implemented a RBM based DBN but using the NSL-KDD dataset. The authors claimed high accuracy up to 97.5\%, but the model was evaluated on 40\% of the holdout training dataset. 
Tang et al. \cite{tang2016deep} proposed a deep learning approach for flow-based anomaly detection in an Software Defined Networking (SDN) environment. The authors  developed a Deep Neural Network (DNN) with 3 hidden layers, trained on the NSL-KDD dataset to perform only  binary classification (normal, anomaly) using six basic features. However, the reported accuracy was  75.75\%. Kim et al. \cite{kim2017method} developed a different DNN architecture (with 4 hidden layers and 100 hidden units) and optimized trained using adam optimization algorithm. However, the  performance was measured using the KDD99 dataset.  Javaid et al. \cite{javaid2016deep} proposed a self-taught learning (STL) approach based on sparse autoencoders for anomaly detection. The NSL-KDD dataset was employed as benchmark to quantify the performance. In  \cite{yin2017deep} proposed a Recurrent Neural Network (RNN) for anomaly detection using the same benchmark, claiming accuracies of 83.28\% and 81.29\% in binary and multiclass classification, respectively. Shone et al. \cite{shone2018deep} proposed a non-symmetric deep auto-encoder (NDAE) model for intrusion detection tested on both KDD99 and NSL-KDD dataset, achieving 5-class  accuracy rate up to 97.85\%  and 85.42\%, respectively. Recently, Diro et al. \cite{abeshu2018deep} proposed a novel DL architecture based on autoencoders for attack detection in fog-to-things computing, using the NSL-KDD. However, the evaluation restricted to binary detection (normal, anomaly) only. In this paper, we propose an innovative statistical driven deep learning method for detecting network intrusion. The NSL-KDD dataset is used to estimate the reliability of the model for binary and multi-class classification and aforementioned limitations have been addressed.
\label{sec:Related works}
\vspace{-1em}
\section{Methodology}
\label{sec:Methodology}
\begin{figure*}[h]
\centerline{\includegraphics[width=12.5 cm]{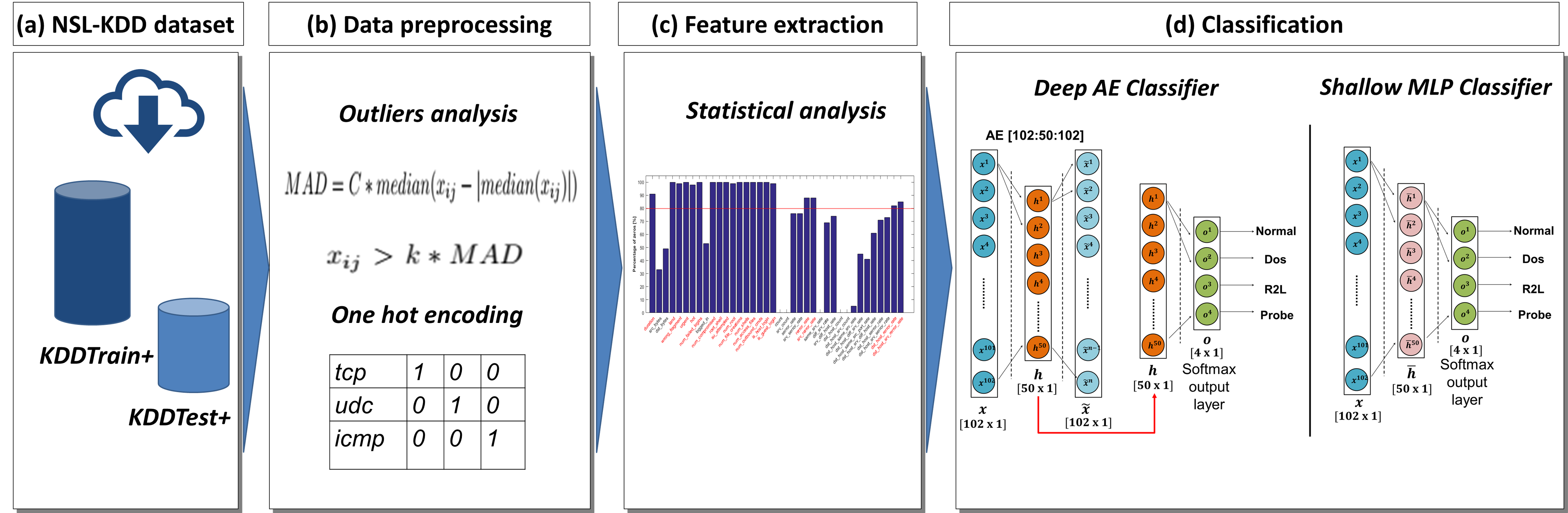}}
\caption{Flowchart of the proposed method}
\label{fig:KDD_method}
\end{figure*}

The flowchart of the proposed method is illustrated in Figure \ref{fig:KDD_method}. The KDD dataset is first filtered by eliminating the outliers, followed by the transformation of  categorical features into  \textit{one-hot-encoded vectors}. Afterwards a statistical analysis has been carried out over 38 numeric features in order to extract more correlated features. Finally, a deep and shallow classifiers are employed to test the detection accuracy. 
\vspace{-1em}
\subsection{NSL-KDD dataset}

The used NSL-KDD dataset \cite{tavallaee2012nsl} is arranged into a training set of 125973 samples (KDDTrain+) and a testing set of 22544 samples (KDDTest+). The dataset has $x_{i}$ (i=1,2,..41) features with  38 numeric and 3 categorical features.  Specifically,  \textit{protocol\_type},  \textit{service},  \textit{flag} ($x_{2}$, $x_{3}$, $x_{4}$) represent 3 categorical variables.  Table \ref{table1} summarizes attack types and 4 different categories: (1)DoS (Denial of Service attacks), (2) R2L (Root to Local attacks), (3) U2R (User to Root attack), (4) Probe (Probing attacks). The structure of the NSL-KDD dataset is shown in Table \ref{table2}.

\vspace{-2em}
%%% start TABLE1%%%%
\begin{table}[h]
\centering
\caption{Attack types of DoS, R2L, U2R, Probe categories. }
\label{table1}
\begin{tabular}{|c|c|}
\hline
\textbf{Attack class} & \textbf{Attack type}                                                                                                     \\ \hline
Dos                   & back, land, neptune, pod, smurf, teardrop                                                                                \\ \hline
R2L                   & \begin{tabular}[c]{@{}c@{}}ftp\_write, guess\_passwd, imap, multihop, \\ phf, spy, warezclient, warezmaster\end{tabular} \\ \hline
U2R                   & buffer\_overflow, loadmodule, perl, rootkit                                                                              \\ \hline
Probe                 & ipsweep, nmap, portsweep, satan                                                                                          \\ \hline
\end{tabular}
\end{table}
\vspace{-2em}
%%% start TABLE 2%%%%
\begin{table}[h]
\centering
\caption{NSL-KDD dataset composition without additional test attacks types.}
\label{table2}
\begin{tabular}{|c|c|c|c|c|c|c|}
\hline
\textbf{NSL-KDD} & \textbf{Total} & \textbf{Normal} & \textbf{Dos} & \textbf{Probe} & \textbf{R2L} & \textbf{U2R} \\ \hline
\textit{KDDTrain+}        & 125973         & 67343           & 45927        & 11656          & 995          & 52           \\ \hline
\textit{KDDTest+}         & 18793          & 9710            & 5741         & 1106           & 2199         & 37           \\ \hline
\end{tabular}
\end{table}
%%% end TABLE 2%%%%

%\vspace{-2em}

\subsection{Data preprocessing}
\subsubsection{Outliers analysis:}Removing outliers from the dataset before performing the data normalization has proven to be an essential task. This operation removes inconsistent values that makes the learning difficult. In this study, the Median Absolute Deviation (MAD) estimator is used for detecting the outliers. It is defined as follow:

\begin{equation}
MAD=C*median(x_{ij}- |median(x_{ij})|)
\end{equation}

\noindent where \textit{C}=1.4826 is a multiplicative constant typically used under the assumption of data normality and $x_{ij}$ is the instance belonged to the feature $x_{i}$.

\noindent Specifically, $x_{ij}$ was considered an outlier when $x_{ij}$ $>$ $k* MAD$ (with k=10). The original size of training and testing sets was reduced from 125973 to 85421 and from 22544 to 11925 , respectively. Table \ref{table3}  summarizes the dataset after removing outliers. It is to be noted that, the dataset is highly unbalanced with only 18 test samples in the U2R attack class. Therefore, the final dataset removed U2R class.

%%% start TABLE 3%%%%
\begin{table}[h]
\centering
\caption{NSL-KDD$^*$ dataset composition after removing outliers.}
\label{table3}
\begin{tabular}{|c|c|c|c|c|c|c|}
\hline
\textbf{NSL-KDD$^*$} & \textbf{Total} & \textbf{Normal} & \textbf{Dos} & \textbf{Probe} & \textbf{R2L} & \textbf{U2R} \\ \hline
${KDDTrain+}^*$        & 85421          & 51551           & 23272        & 9683           & 874          & 41           \\ \hline
${KDDTest+}^*$         & 11925          & 7341            & 1975         & 620           &1971           & 18           \\ \hline
\end{tabular}
\end{table}
%%% end TABLE 3%%%%
\vspace{-0.7em}
\subsubsection{One hot encoding} Since features ${x}_{2}$, ${x}_{3}$, ${x}_{4}$ (\textit{protocol type}, \textit{service} and \textit{flag}) consist of categorical values, these features were converted into \textit{one hot encoded vectors}. For example, the \textit{protocol type} feature includes 3 attributes: \textit{tcp}, \textit{udp} and \textit{icmp}, and were represented as (1,0,0),(0,1,0),(0,0,1), respectively. Similarly, \textit{service} and \textit{flag} features were represented by binary values. 
This procedure maps the 41-dimensional features into 122-dimensional features: 38 continuous and 84 with binary values associated to the 3 categorical features ( \textit{protocol}, \textit{flag}, and \textit{service}). 
\label{sssec:Data preprocessing}
\subsection{Feature extraction}
\begin{figure*}[h]
\centerline{\includegraphics[width=12 cm]{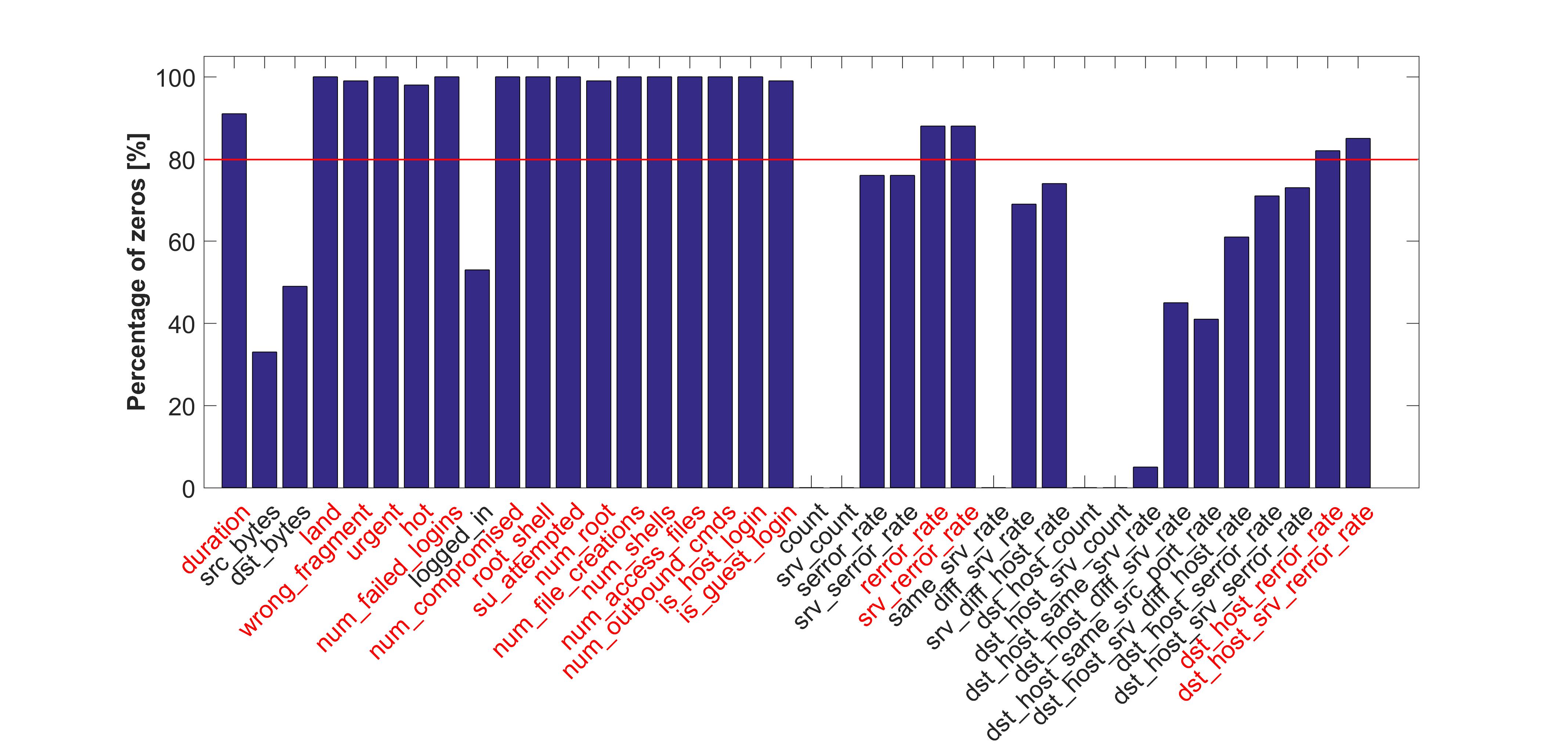}}
\caption{Distribution of the number of zeros in each numeric features of the training set. Features with null values greater than 80\% are depicted in red and are discarded from the analysis.}
\label{fig:bar_plot_zeros}
\end{figure*}

Several feature extraction techniques exist in literature (i.e.\cite{morabito2000independent}). However, in this study, the proposed feature extraction module selects the most correlated features according to the following procedure. Firstly, the percentage of null values are quantified for 38 numeric features vectors. The histogram in Figure \ref{fig:bar_plot_zeros} shows the distribution of zeros of each numeric feature in the training set. In this study, variables with number of zeros greater than 80 \% are not included in the analyses.It is to be noted that  20 features (depicted in red in Figure \ref{fig:bar_plot_zeros} ) out of 38 consists of mostly zeros which are discarded. Therefore, the remaining 18 numeric features are combined with 84 one-hot-encoded features forming 102-dimensional input vector for deep AE and shallow MLP classifier.
\subsection{Classification}
In this section, two machine learning algorithms are presented to classify different categories of the NSL-KDD dataset (Normal, DoS, R2L and Probe). Specifically, a deep architecture based on Autoencoder and a shallow architecture based on standard Multi Layer Perceptron are implemented. Details of the proposed frameworks are described in the following subsections.
\subsubsection{Deep AE Classifier}
An autoencoder is a type of unsupervised learning algorithm, typically used for dimensionality reduction purposes. The AE standard configuration includes one input layer, one output layer and one hidden layer, as showed in Figure \ref{fig:AE_MLP_classifiers} (a). It compresses the input data \textit{x} into a lower dimension \textit{h} through the encoding process:
\begin{equation}
h=g(xw+b)
\end{equation}
\noindent where \textit{x}, \textit{w}, \textit{b} are the input vector, weight matrix, the bias vector, respectively and \textit{g} is the activation function.
\noindent Then, it attempts to reconstruct the same set of input (\textit{x}) from the compressed representation (\textit{h}) through the decoding process:
\begin{equation}
\tilde{x}=g(hw^{T}+b)
\end{equation}
 
\noindent The architecture of the deep AE classifier is showed in Figure \ref{fig:AE_MLP_classifiers} (a). The extracted 102 features are the input of the single hidden layer AE that compressed the input space from 102 into 50 latent features ({$AE_{[50]}$).
At this stage, the AE is trained with unsupervised learning through the scaled conjugate gradient algorithm, for 100 iterations.  The \textit{saturating linear transfer function} ($g(z)=0$ if $z\leq{0}$, $g(z)=z$ if $0<z<1$, $g(z)=0$ if $z\geq{1}$) and the \textit{linear transfer function} ($g(z)=z$) is used for encoding and decoding operations. 
The reconstruction of the input features (\textit{x}) is measured through the mean squared error (MSE) coefficient. The proposed AE achieved  the reconstruction error of 0.0083.  Subsequently, the 50 compressed features are fed into a softmax output layer trained with supervised learning for performing the multi-class detection task. Finally, the whole network (AE+softmax) is trained with supervised learning (backpropagation algorithm) to improve the classification performance (fine-tuning method). The training is stopped when the cross-entropy loss function \cite{de2005tutorial} saturates. In this study, the convergence is observed after 300 iterations.

\vspace{-2em}
\subsubsection{Shallow MLP Classifier}
\noindent The architecture of the shallow MLP classifier is showed in Figure \ref{fig:AE_MLP_classifiers} (b).  it constitutes standard  standard feed-forward neural network trained  with supervising learning through scaled conjugate gradient algorithm. For fair comparative analysis, both shallow and deep AE models have adopted same architecture. The shallow MLP architecture has used a single hidden layer with 50 hidden neurons followed by a softmax output layer.

\begin{figure*}[h]
\centerline{\includegraphics[width=10.5 cm]{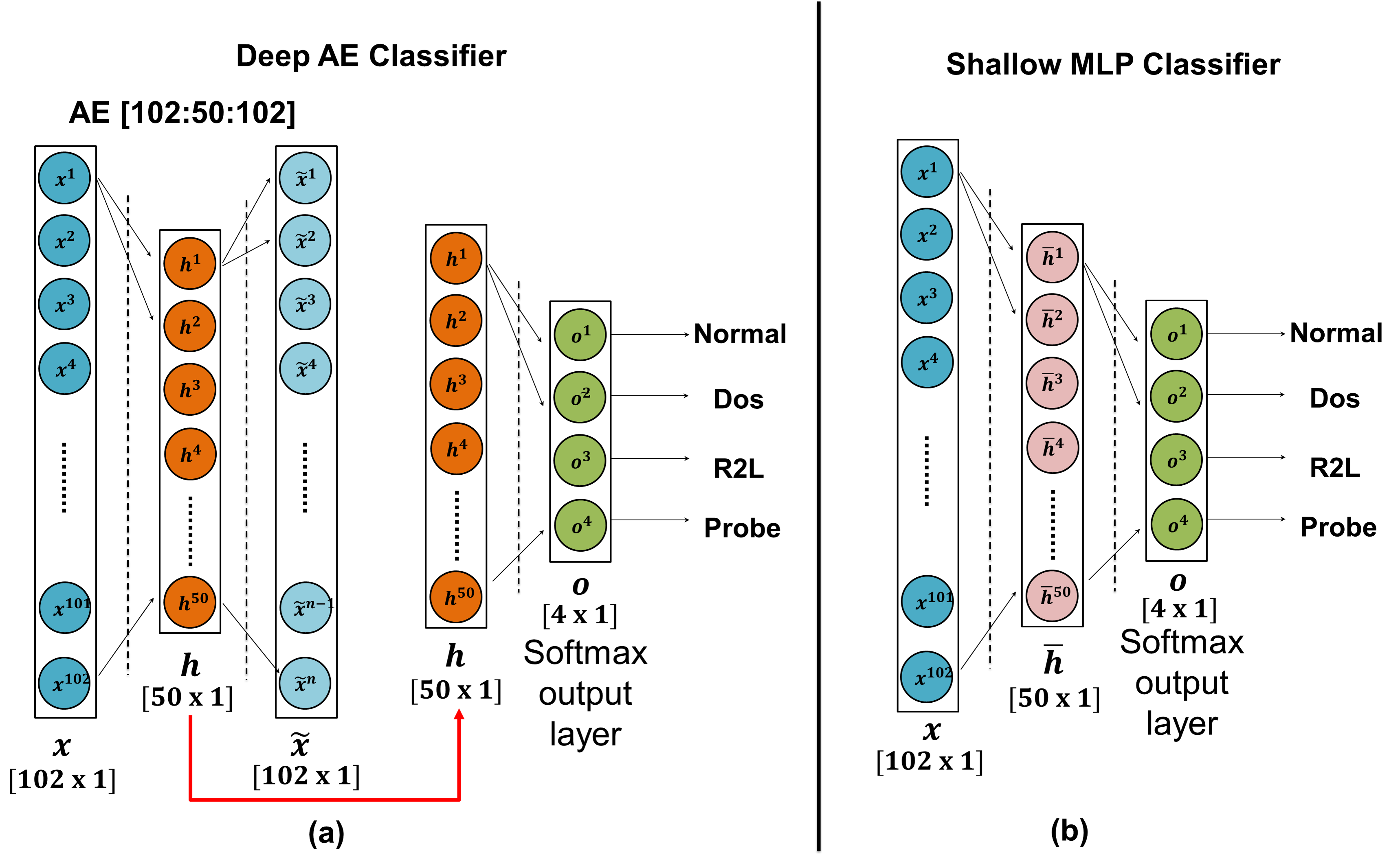}}
\caption{(a) Proposed Deep AE Classifier: The AE [102:50:102] compresses the 102 features (\textbf{\textit{x}}) into 50  most significant variables (\textbf{\textit{h}}) used as input for a final softmax output layer (\textbf{\textit{o}}) to perform the multi-class detection. (b) Shallow MLP Classifier: standard single layer feed-forward neural network with 50 hidden neurons (\textbf{\textit{$\tilde{h}$}}) followed by a softmax output layer (\textbf{\textit{o}}) .  }
\label{fig:AE_MLP_classifiers}
\end{figure*}
\label{sssec:Classification}

%\vspace{-2em}

\section{Experimental results}
The effectiveness of the proposed deep (AE) and shallow (MLP) classifier is evaluated using standard metrics: Precision (PR), recall (RC), F\_measure (or F\_score) and accuracy (ACC):
\begin{equation}
Precision =\frac{TP}{TP+FP}
\end{equation}
\begin{equation}
Recall =\frac{TP}{TP+FN}
\end{equation}
\begin{equation}
F\_measure =2*\frac{Precision*Recall}{Precision+Recall}
\end{equation}
\begin{equation}
Accuracy =\frac{TP+TN}{TP+TN+FP+FN}
\end{equation}
where true positives (TP) represent the number of anomalous samples correctly identified as anomaly; true negatives (TN) represent the number of normal samples correctly identified as normal; false positives (FP) are the number of normal samples missclassified as anomaly; false negatives (FN) are the number of anomaly samples missclassified as normal.  Table \ref{table4} reports the outcome of the experiments. The shallow MLP classifier showed good performances in detecting Normal, Dos and Probe categories, reporting F\_measure values of 87.10\%, 97.08\%, 77.13\%, respectively. However remained deficient  to classify the R2L attack class accurately ( achieving F\_measure of 11.74\%).The  deep  AE  classifier  achieved  better  results  in  terms of F\_measure rate in all categories, with performances up to 98\%. To find out the most optimal AE architecture, different numbers of hidden layers were tested. Specifically, the performance of the proposed deep AE classifier (having one layer) with 50 hidden neurons ($AE_{[50]}$) was compared with a two hidden layers AE architecture (with 50 and 25 hidden neurons respectively, $AE_{[50,25]}$) and a three hidden layers AE architecture (with 50, 25, 12 hidden neurons respectively, $AE_{[50,25,12]}$). Experimental results showed that the $AE_{[50]}$ classifier achieved the best accuracy of 87\%  as compared to accuracies achieved by $AE_{[50,25]}$ (82\%) and $AE_{[50,25,12]}$ (81\%), this is possibly due to over compression in multilayered AE.
\vspace{-1em}
%%%%% start Table4 %%%%%%%
\begin{table}[h]
\centering
\caption{NSL-KDD$^*$ performance (Precision, Recall, F\_measure) for the deep AE classifier and the shallow MLP classifier.}
\label{table4}
\begin{tabular}{|c|c|c|c|c|c|c|}
\hline
{\textbf{Attack class}} & \multicolumn{2}{c|}{\textbf{Precision}} & \multicolumn{2}{c|}{\textbf{Recall}} & \multicolumn{2}{c|}{\textbf{F\_measure}} \\ \cline{2-7} 
                                       & \textbf{MLP}        & \textbf{AE}       & \textbf{MLP}      & \textbf{AE}      & \textbf{MLP}        & \textbf{AE}        \\ \hline
Normal                                 & 96.35               & 96.19             & 79.46             & 85.03            & 87.10               & 90,27              \\ \hline
Dos                                    & 96.96               & 98.18             & 97.21             & 97.05            & 97.08               & 97,61              \\ \hline
Probe                                    & 96.29               & 94.03             & 64.33             & 69.82            & 77.13               & 80,14              \\ \hline
R2L                                & 6.24                & 39.78             & 99.19             & 99.49            & 11.74               & 56,83              \\ \hline
\end{tabular}
\end{table}
%%%%% end Table4 %%%%%%%

\noindent Table \ref{table5} reports the comparison of proposed deep AE and shallow MLP classifier with four recently proposed approaches in the literature.
In order to compare our results, we referred to studies that focused on multi-classification tasks using the NSL-KDD dataset. It is to be noted that, the proposed AE based deep learning architecture, outperformed all other approaches, achieving accuracy up to 87\%. However, the MLP classifier achieved maximum accuracy of 81.6\% . In \cite{huang2017work} the authors proposed a sequential learning algorithm achieving an accuracy of 76.04\%. In \cite{javaid2016deep} the authors developed a sparse autoencoder based classifier and reported 79.10\% accuracy.  In \cite{yin2017deep} the authors developed a RNN based system and reported multiclass accuracy of 81.29\%. In \cite{shone2018deep}, the authors proposed a stacked non-symmetric deep autoencoders and reported 5-class accuracy of 85.42\%. Here, instead, we propose an alternative statistical analysis driven deep AE classifier able to  achieve multiclass accuracy of 87\%.
%\vspace{-2em}

%%%%% start Table 5 %%%%%%%
\begin{table}[h]
\centering
\caption{Accuracy performance and comparison with state of the art models.}
\label{table5}
\begin{tabular}{|c|c|}
\hline
\textbf{Model} & \textbf{Accuracy (\%)} \\ \hline
\textbf{AE proposed}    & \textbf{87}                     \\ \hline
MLP proposed   & 81.43                   \\ \hline
Huang et al. \cite{huang2017work}   & 76.04                  \\ \hline
Abeshu et al. \cite{javaid2016deep} & 79.10                  \\ \hline
Yin et al. \cite{yin2017deep}    & 81.29                  \\ \hline
Shone et al. \cite{shone2018deep}   & 85.42                  \\ \hline
\end{tabular}
\end{table}

%%%%% end Table 5%%%%%%

\vspace{-1em}
\label{sec:Experimental results}
\section{Conclusion}
The proposed approach leverages the complementary strengths of both automated feature engineering (provided by deep learning) and manual statistical driven optimized feature engineering (based on human-in-the-loop and big data visualization) that helps the learning model to better correlate the input-output relationship.
Specifically, we introduced a statistical analysis driven optimized DL system for intrusion detection. The NSL-KDD dataset was employed as benchmark to identify normal and abnormal network traffic patterns.
The most correlated features were extracted using statistical methods and were the input of a deep AE classifier. The feasibility and effectiveness of the proposed model were evaluated using precision, recall, F\_measure and accuracy metrics.
The comparative evaluation of proposed deep autoencoder with a shallow MLP classifier and state-of-the-art models showed that the proposed deep AE classifier outperformed all other approaches and achieved upto 87\% accuracy.
\noindent Future works include a more robust system capable of handling real-time traffic similar to NSL-KDD dataset to contextually detect intrusions in real-time applications. In addition, to concurrently acquire long-term learning, fast decision making, and low computational complexity for real-time Big Data processing, we intend to integrate the proposed work with our previously presented real-time autonomous decision making system \cite{adeel2016random}.  
%Further works include  
\label{sec:Conclusion}
\vspace{-1em}
\section{Acknowledgment}
Amir Hussain and Ahsan Adeel were supported by the UK Engineering and Physical Sciences Research Council (EPSRC) grant No.EP/M026981/1.
\vspace{-1em}
 \bibliographystyle{splncs04}
 \bibliography{cyberbib}
%Author, F.: Article title. Journal \textbf{2}(5), 99--110 (2016)
%Author, F., Author, S.: Title of a proceedings paper. In: Editor,
%F., Editor, S. (eds.) CONFERENCE 2016, LNCS, vol. 9999, pp. 1--13.
%Springer, Heidelberg (2016). \doi{10.10007/1234567890}
%Author, F., Author, S., Author, T.: Book title. 2nd edn. Publisher,
%Author, A.-B.: Contribution title. In: 9th International Proceedings
%on Proceedings, pp. 1--2. Publisher, Location (2010)
%LNCS Homepage, \url{http://www.springer.com/lncs}. Last accessed 4
%\end{thebibliography}
\end{document}